\documentclass{elsart}

\usepackage{natbib}

\usepackage{graphicx}

\usepackage{amssymb}

\begin{document}

\begin{frontmatter}



\title{Epitaxy of thin films of the Heusler compound Co$_2$Cr$_{0.6}$Fe$_{0.4}$Al}


\author{A.\,Conca}
\author{M.\,Jourdan}
\ead{jourdan@uni-mainz.de}
\author{C.\,Herbort}
\author{H.\,Adrian}
\address{Institute of Physics, Johannes Gutenberg University Mainz, Staudingerweg 7, 55128 Mainz, Germany}

\begin{abstract}
Epitaxial thin films of the highly spin polarized Heusler compound Co$_2$Cr$_{0.6}$Fe$_{0.4}$Al are deposited
by dc magnetron sputtering. It is shown by XRD and TEM investigations how the use of an Fe buffer layer on
MgO(100) substrates supports the growth of highly ordered Co$_2$Cr$_{0.6}$Fe$_{0.4}$Al at low deposition 
temperatures. The as grown samples show a relatively large ordered magnetic moment of $\mu \simeq 3.0\mu_B/f.u.$
providing evidence for a low level of disorder.
\end{abstract}


\end{frontmatter}

\section{\label{sec:level1}Introduction}
The compound Co$_2$CrAl belongs to the Heusler-type (L2$_1$ structure) materials for which
100 \% spin polarization at the Fermi energy, i.\,e.\,half metallic properties are
predicted \cite{Gal02a,Fec05}. For potential technical applications the magnetic
ordering temperature of this material has to be raised well above room temperature
by doping with iron \cite{Blo03}. However, considering real samples the effects of
impurities and crystal imperfections have to be taken into account \cite{Miu04, Fec05}.
Additionally, at surfaces and interfaces the local spin polarization depends as well 
on the crystallographic surface orientation as on the properties of the interface 
partner material \cite{Gal02, Nag04}.\\
For those reasons it is important to grow single crystalline thin films with low defect
concentration which can be used for various investigations like spin polarized photo 
electron spectroscopy or the integration in magnetic tunnel junctions. Polycrystalline 
Co$_2$Cr$_{0.6}$Fe$_{0.4}$Al thin films can be grown on oxidized Si \cite{Ino06}, 
(110)-oriented films were obtained on Al$_2$O$_3$(110) \cite {Jak05} and (100) oriented
epitaxial growth was observed directly on MgO(100) \cite{Ino06, Mat06}.\\
Here we report how the use of an Fe buffer layer on an MgO(100) substrate assists the 
epitaxial growth of high quality Co$_2$Cr$_{0.6}$Fe$_{0.4}$Al thin films deposited at low
temperatures without the need for a high temperature annealing step.

\begin{figure}
\includegraphics[width=0.5\columnwidth,angle=-90]{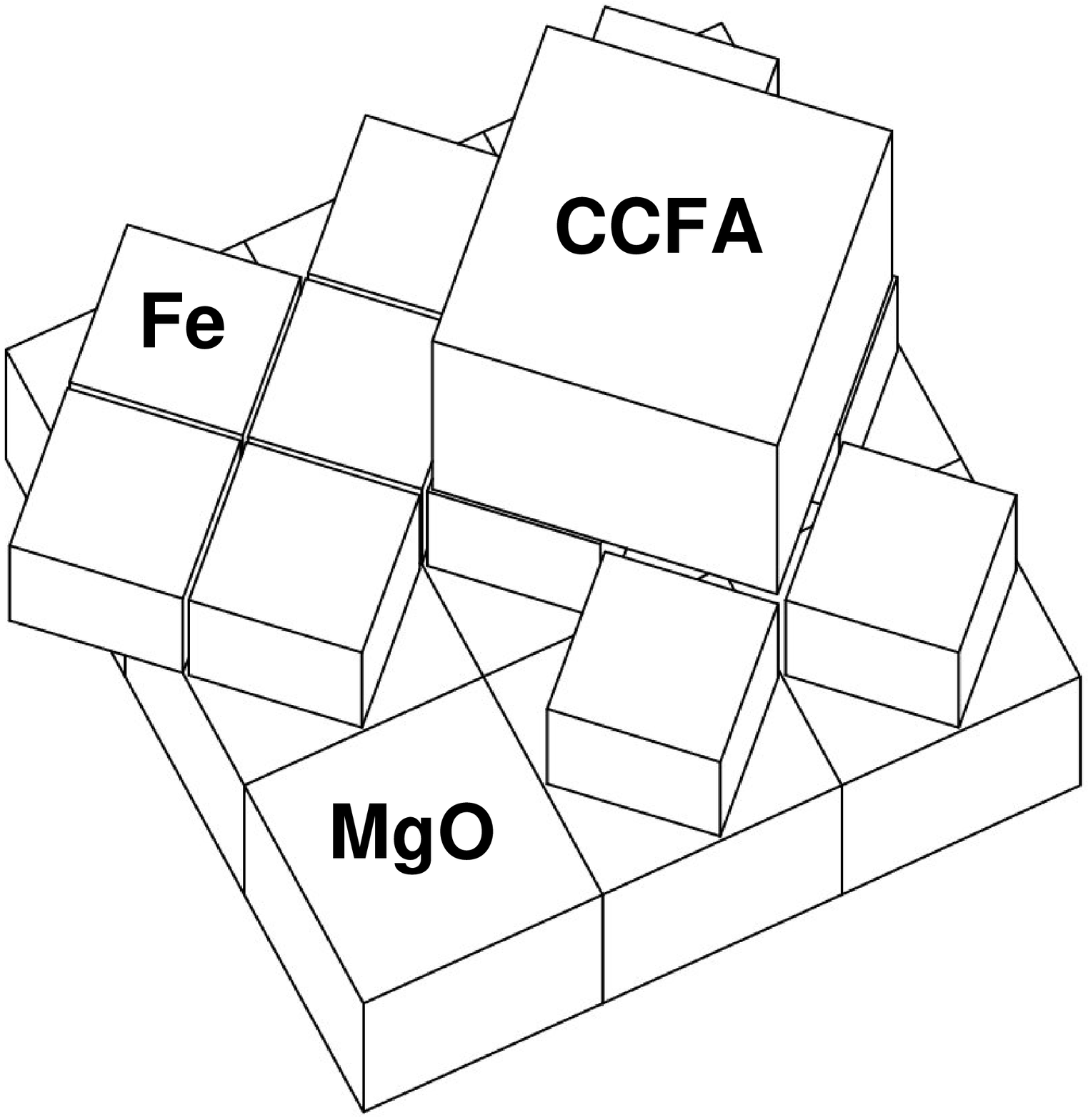}
\caption{\label{epidars} Schematic representation of the epitaxial relationship of a CCFA thin film on an Fe buffer layer on a MgO (100) substrate.}
\end{figure}  

\section{\label{sec:level2} Preparation}
Co$_2$Cr$_{0.6}$Fe$_{0.4}$Al (CCFA) films were deposited by dc magnetron sputtering on MgO(100) substrates. The target stoichiometry as given by the supplier (TBL-Kelpin, Neuhausen) was confirmed by EDX (SEM). By the same method applied in a TEM the thin film stoichiometry was shown to be consistent with the target stoichiometry within the typical experimental error of 10\%.  Before deposition the commercial substrates (Crystec, Berlin) were annealed ex situ in an oxygen atmosphere at 950$^0$C for 2 hours and subsequently exposed to a microwave oxygen plasma. However, a direct deposition at low temperatures (T $\simeq 100^{\rm o}$C) of CCFA on the substrates did not result in the formation of  epitaxial CCFA thin films. Considering the small lattice misfit between Fe and CCFA of $\simeq 0.1\%$, Fe was selected as a buffer layer for the deposition of the Heusler-compound. The expected epitaxial relation of the substrate and thin film layers is shown in Fig.\,\ref{epidars}. An 8nm thick Fe buffer layer was deposited by electron beam evaporation in a separate MBE chamber which is part of the deposition system. The CCFA thin films were prepared in a sputtering chamber with a base pressure of $\simeq 3\times 10^{-8}$mbar on the buffer layer at temperature $T\simeq 100^{\rm o}$C. In p$=6\times 10^{-3}$mbar of Ar a deposition rate of 0.5 nm$/$s was selected. All films were protected against oxidation by a capping layer of 4nm of Al before removing them from the deposition chamber.    
\\
\begin{figure}
\includegraphics[width=0.7\columnwidth,angle=-90]{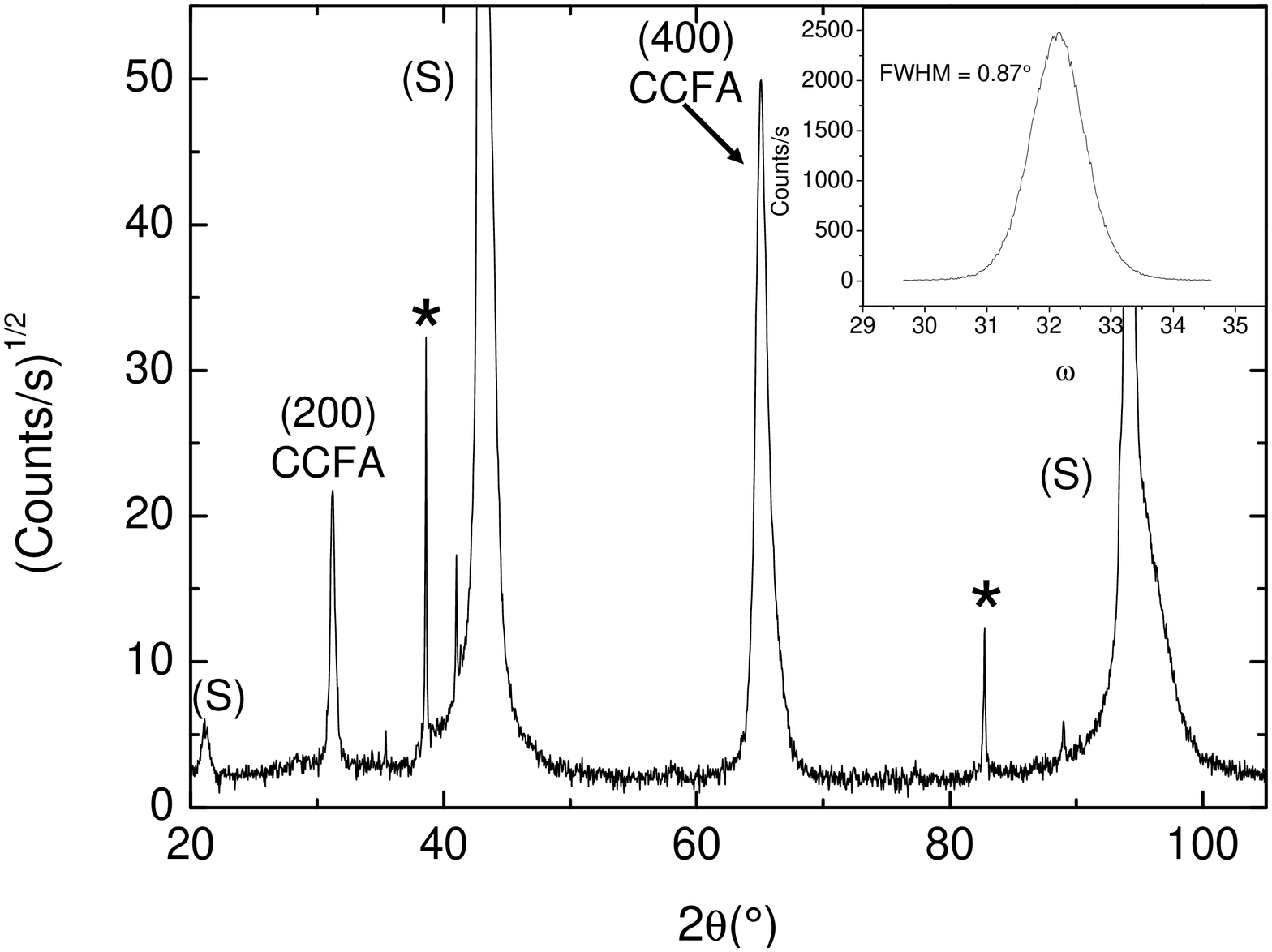}
\caption{\label{diff} X-ray $\Theta$/2$\Theta$-scan (Bragg-Brentano geometry) of a 100 nm thick CCFA film deposited at 100$^{\rm o}$C. The inset shows the rocking curve ($\omega$-scan ) of the (400) reflection. The substrate reflections are marked (S), substrate reflections due to secondary x-rays wavelengths are marked (*).}
\end{figure}

\begin{figure}
\includegraphics[width=0.7\columnwidth,angle=-90]{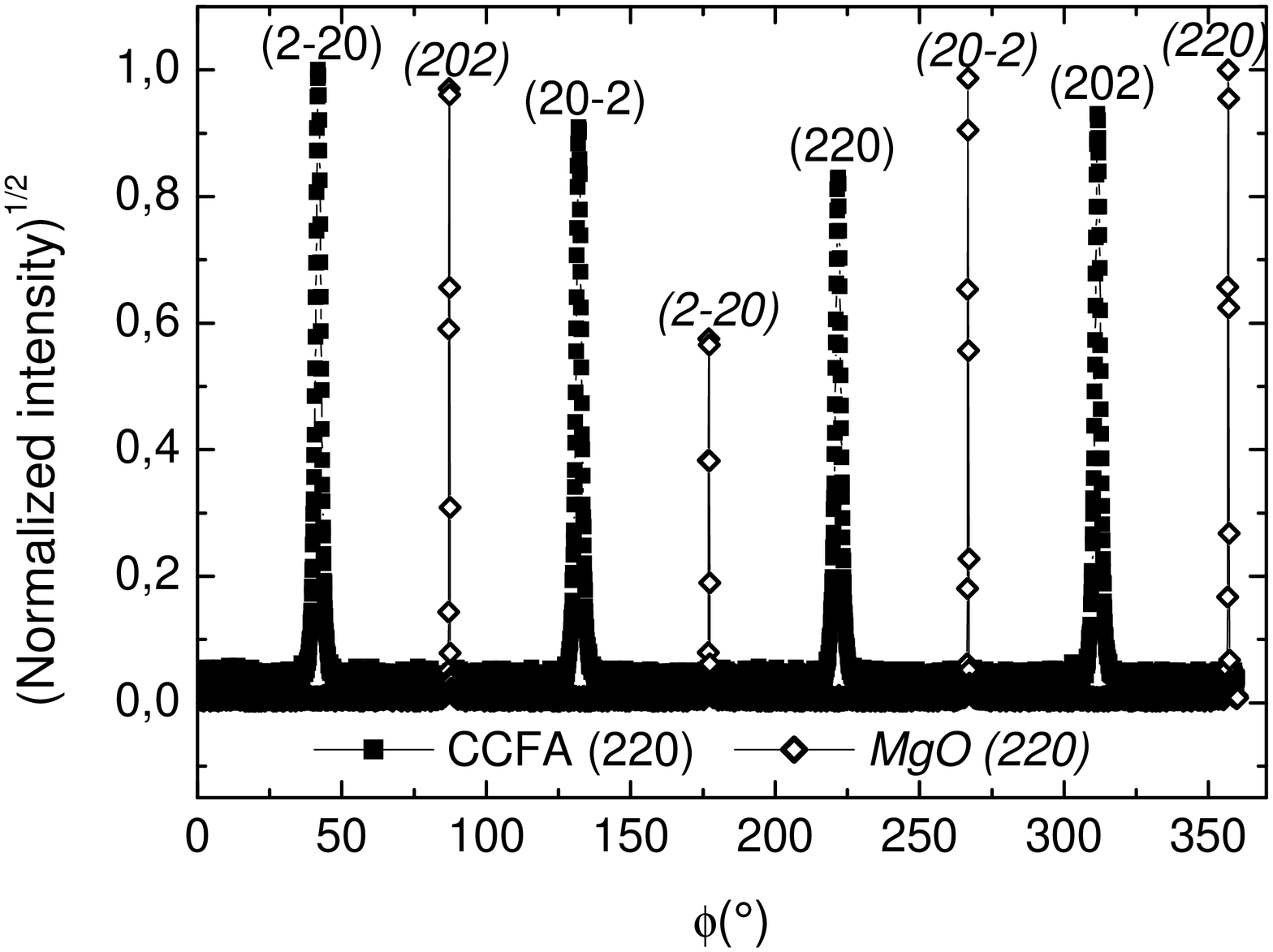}
\caption{\label{phiscan} X-ray $\phi$-scan of the (220) equivalent reflections of a CCFA film deposited at 100$^{\rm o}$C (filled symbols) and of the MgO substrate (open symbols, peaks labeled in italics). The different scattering intensities of equivalent reflections are due to a misalignment of the sample on the goniometer.}
\end{figure}

\section{\label{sec:level3}Crystallographic order and epitaxial relation}
Different types of site disorder are possible in CCFA. Cr-Al disorder (B2 structure) is most likely due to a very small difference in the total formation energy of the ordered and disordered structure \cite{Miu04}. However, this type of disorder is predicted to have only a small influence on the spin polarization of CCFA. Possible types of disorder which strongly reduce the spin polarization are Co-Cr disorder or disorder on all sites (A2 structure) \cite{Miu04}.\\   
For the investigation of the crystallographic properties of the thin films x-ray diffraction was employed. The characteristic x-ray reflections for the disorder in CCFA are (111) and (200). Both are present in the cases of the fully ordered L2$_1$ structure and for pure Co-Cr disorder (with different intensity ratio). However, (111) disappears completely if there is full disorder on the Cr-Al positions. If there is disorder on all positions (A2 structure) (200) disappears as well.\\ 
Fig.\,\ref{diff} shows a $\Theta/2\Theta$-scan of a CCFA thin film obtained in Bragg-Brentano geometry in which scattering at Bragg-planes which are parallel to the substrate surface is observed (specular reflections). The (200) reflection of CCFA is clearly visible. The stronger (400) reflection covers the same angle as the Fe (200) peak of the thin buffer layer (Fe (100) is symmetry forbidden). The observation of the CCFA (200) reflection excludes already the A2 structure, but further insight is obtained from 4-circle x-ray diffraction which allows the investigation of off-specular reflections.\\
In  Fig.\,\ref{phiscan} a $\phi$-scan of the (220) equivalent reflections of a film and substrate is shown. In this scan the film normal is tilted by $45^{\rm o}$ out of the scattering plane and the sample is rotated by the angle $\phi$ around the film normal. The scan shows that the  film is in-plane ordered and proves the epitaxial growth. From the observation of the peak positions of CCFA which are rotated by 45$^{\rm o}$ with respect to the MgO substrate peak positions it is concluded that the epitaxial relation is indeed as shown in Fig.\,\ref{epidars}.\\
Another off-specular peak, the  (111) reflection, was not observed in our films. This, in combination with the observation of the (200) reflection, indicates that there is full disorder on the Cr-Al positions, but order on the Co positions. From the ratio of the scattering intensities of (200) and (400) and considering a geometrical correction due to the thin film geometry of the sample a lower limit for remaining disorder on the Co sites can be estimated:  From (I$_{(200)}\times sin\Theta_{(200)}$)/(I$_{(400)}\times sin\Theta_{(400)}$)$\simeq0.16$ and comparing to a simulation (PowderCell) it can be concluded that less than 18\% of the Co sites are occupied by other atoms.  Thus the films grow with a high degree of B2 order.

\section{\label{sec:level4} Microstructural properties}

\begin{figure}
\includegraphics[width=0.7\columnwidth,angle=0]{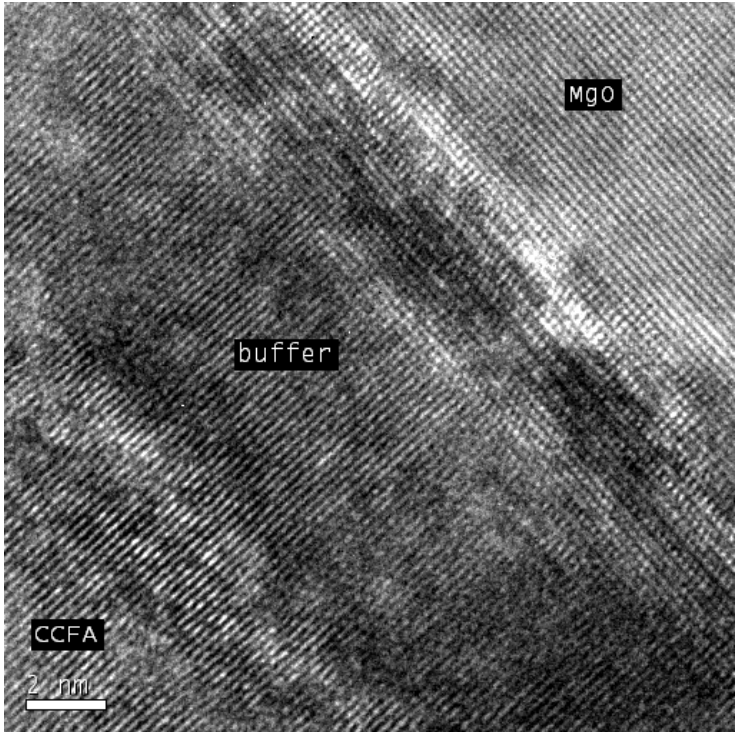}
\caption{\label{ch_buffer} HRTEM image in (010) direction of CCFA of the interface region with MgO substrate, Fe buffer layer and CCFA thin film.}
\end{figure}

For a direct access to the substrate-film interface and in particular to reveal the function of the buffer layer the samples were investigated by high-resolution transmission electron microscopy (HRTEM) at the center for electron microscopy of the university of Mainz (EMZM). The TEM cross sections were prepared by mechanical thinning and argon-ion polishing.\\
Fig.\,\ref{ch_buffer} shows an image in (010) direction of CCFA of the interface region with MgO substrate, Fe buffer and CCFA thin film.

At the interface between MgO and Fe some lattice distortions are visible, which can be related to the lattice mismatch of $(a_{2 Fe}-\sqrt{2} a_{MgO})/a_{2 Fe}\simeq-0,037$. However, these distortions disappear after some atomic layers and the structure of the buffer layer is well ordered. The interface between Fe and CCFA can not be clearly identified from the HRTEM images. No distortion of the atomic layers is observable at this interface. This indicates a perfect epitaxial growth of CCFA on the buffer layer due to the very small lattice mismatch of $(a_{CCFA}-a_{2 Fe})/a_{CCFA}\simeq-0,001$.

\section{\label{sec:level5} Magnetic properties}

The magnitude of the magnetic moment per formula unit (f.u.) of CCFA is an important figure of merit for the film quality. Typical values reported for CCFA thin films prepared including a high temperature annealing process amount to ${\rm \mu_{CCFA}\simeq 2.7 \mu_B/f.u.}$ \cite{Mat06}.\\

\begin{figure}
\includegraphics[width=0.7\columnwidth,angle=-90]{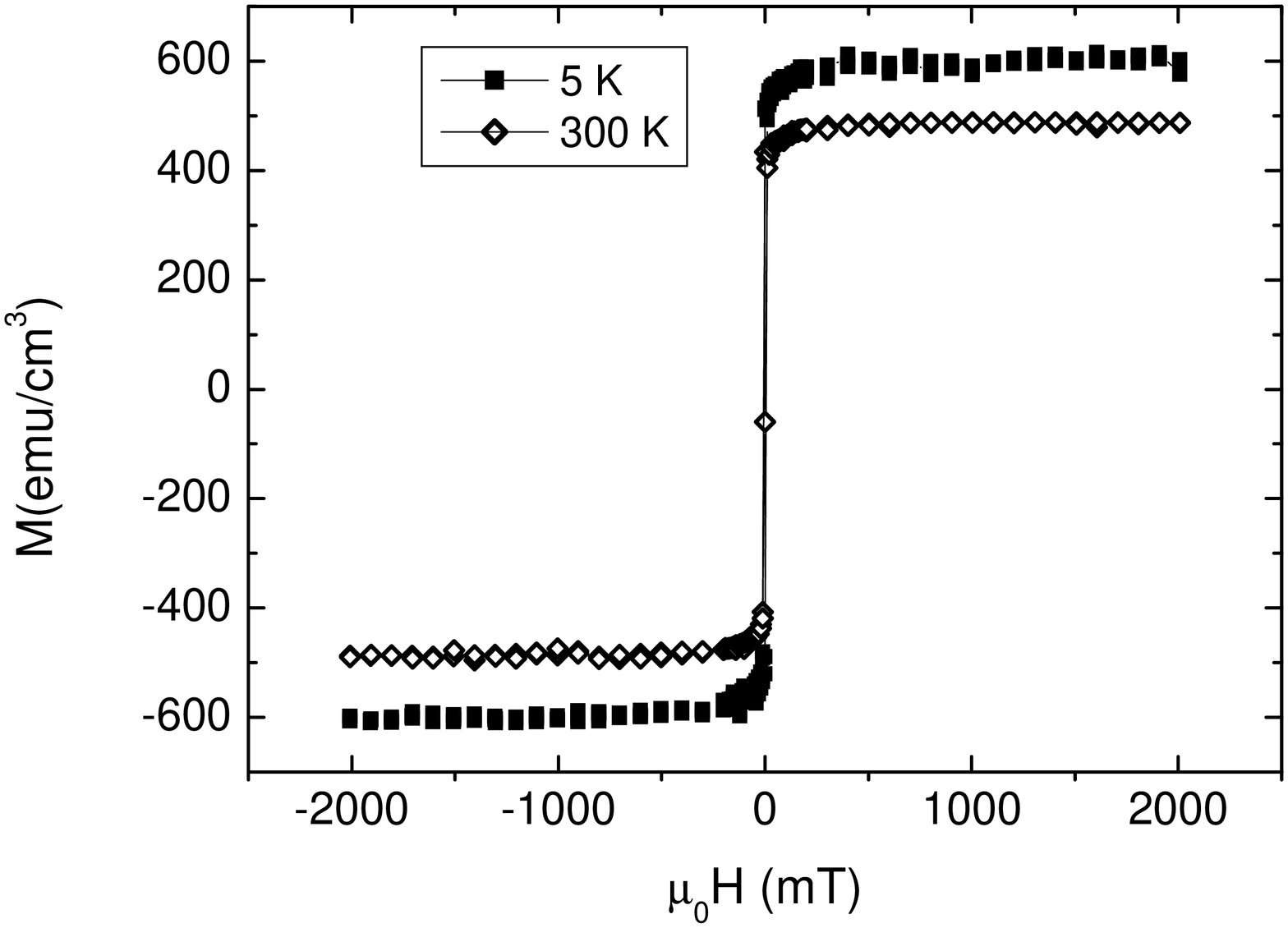}
\caption{\label{hyst} Hysteresis curves of a 100 nm thick CCFA film deposited on Fe at 100$^{\rm o}$C measured at T = 5K (solid squares) and T = 300K (empty diamonds). The contribution of the substrate and buffer layer to the total magnetization is subtracted.}
\end{figure}

\begin{figure}
\includegraphics[width=0.7\columnwidth,angle=-90]{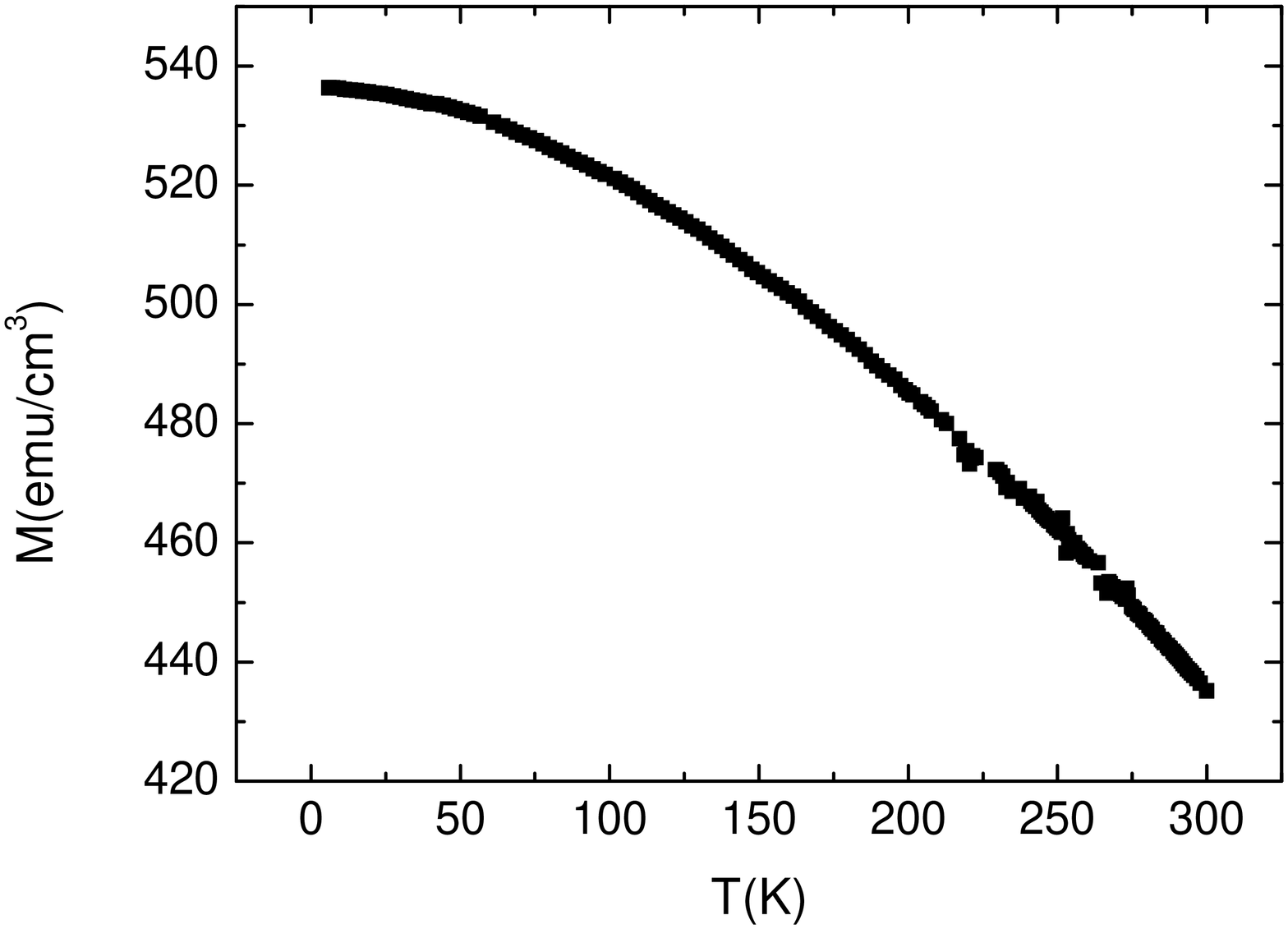}
\caption{\label{mvont} Temperature dependence of the magnetization of a 100 nm thick CCFA film deposited on Fe at 100$^{\rm o}$C. Remanence measured in a magnetic field ${\rm \mu_0H = 20mT}$. The contribution of the substrate and buffer layer to the total magnetization is subtracted.}
\end{figure}

The magnetic properties of our thin films were analyzed using a SQUID magnetometer (Quantum Design MPMS). The contributions of the MgO substrate and the Fe buffer layer were measured separately and were subtracted from the total magnetization of the complete sample.  In Fig.\ \ref{hyst} the hysteresis curves measured at 5K and 300K are shown. The measured volume magnetization corresponds to ${\rm \mu_{CCFA}\simeq 3.0 \mu_B/f.u.}$ at T = 5K. The deviation from the theoretically predicted value for CCFA, 3.8 $\mu_B$ \cite{Gal02a}, may be explained by a relatively small partial disorder on the Co-Cr atomic positions \cite{Miu04}.\\
The temperature dependence of the magnetic moment of the CCFA thin films is shown in Fig.\ \ref{mvont}. At T = 300K the magnetic moment amounts to ${\rm \mu_{CCFA}\simeq 2.5 \mu_B/f.u.}$. This reduction of the magnetic moment compared to the low temperature value is more pronounced than the dependence theoretically predicted for fully ordered Co$_2$CrAl \cite{Sas05}.

\section{\label{sec:level6} Summary}
Epitaxial (100) oriented thin films of the Heusler compound Co$_2$Cr$_{0.6}$Fe$_{0.4}$Al (CCFA) were grown by dc magnetron sputtering on MgO(100) substrates employing an Fe (100) buffer layer. This buffer layer is responsible for the high crystallographic quality of the CCFA thin films by relaxing the strain due to the lattice misfit with the substrate and providing an ideal seed for distortion free epitaxial growth of the Heusler compound.
A relatively large magnetic moment per formula unit of $\simeq 3.0 \mu_{\rm B}$ was observed indicating only a small degree of disorder on the Co-Cr sites.\\
Due to the Fe buffer layer CCFA thin films with the B2 structure can be obtained at relatively low substrate deposition temperatures T = $100{\rm ^o}$C without the need for an additional high temperature annealing process. This may be helpful considering technological applications of CCFA, e.\ g.\ the integration into magnetic tunneling junctions or spin valves.\\

{\bf acknowledgments}\\
This project is financially supported by the {\em Stiftung Rheinland-Pfalz f\"ur Innovation}. Experimental support by F.\,Banhart concerning the HRTEM investigations is gratefully acknowledged.\\

\end{document}